\documentclass[twocolumn,apj,appendixfloats,numberedappendix]{emulateapj-rtx4}
\usepackage{epstopdf}

\shorttitle{Resistive MHD simulations of the ``ideal'' tearing mode}
\shortauthors{Landi et al.}

\newcommand{\be}{\begin{equation}}
\newcommand{\ee}{\end{equation}}
\newcommand{\bdm}{\begin{displaymath}}
\newcommand{\edm}{\end{displaymath}}
\newcommand{\bea}{\begin{eqnarray}}
\newcommand{\eea}{\end{eqnarray}}
\newcommand{\ba}{\begin{align}}
\newcommand{\ea}{\end{align}}

\begin{document}

\title{Resistive magnetohydrodynamic simulations of the {\it ideal} tearing mode}

\author{
S. Landi$^{1,2}$,
L. {Del Zanna}$^{1,2,3}$,
E. Papini$^{4}$, 
F. Pucci$^{5}$, and
M. Velli$^{6}$
}

\affil{$^1$Dipartimento di Fisica e Astronomia, Universit\`a degli Studi di Firenze, L.go E. Fermi 2, 50125 Firenze, Italy.}
\affil{$^2$INAF - Osservatorio Astrofisico di Arcetri, L.go E. Fermi 5, 50125 Firenze, Italy.}
\affil{$^3$INFN - Sezione di Firenze, Via G. Sansone 1, 50019 Sesto Fiorentino (Firenze), Italy.}
\affil{$^4$Max-Planck-Institut f\"ur Sonnensystemforschung, J.~von Liebig Weg 3, 37077 G\"ottingen, Germany.}
\affil{$^5$Universit\`a di Roma Tor Vergata, Dipartimento di Fisica, Via della Ricerca Scientifica 1, 00133 Roma, Italy.}
\affil{$^6$Earth Planetary and Space Sciences, University of California, 595 Charles Young Drive East, Los Angeles, CA 90095, USA.}

%
%

\email{simone.landi@unifi.it}

\begin{abstract}
We study the linear and nonlinear evolution of the tearing instability on thin current sheets by means of two-dimensional numerical simulations, within the framework of compressible, resistive magnetohydrodynamics. In particular we analyze the behavior of current sheets whose inverse aspect ratio scales with the Lundquist number $S$ as $S^{-1/3}$. This scaling has been recently recognized to yield the threshold separating fast, \emph{ideal} reconnection, with an evolution and growth which are \emph{independent of $S$} provided this is high enough, as it should be natural having the ideal case as a limit for $S\to\infty$. Our simulations confirm that the tearing instability growth rate can be as fast as $\gamma\approx 0.6\,{\tau_A}^{-1}$, where $\tau_A$ is the ideal Alfv\'enic time set by the macroscopic scales, for our least diffusive case with $S=10^7$. The expected instability dispersion relation and eigenmodes are also retrieved in the linear regime, for the values of $S$ explored here. Moreover, in the nonlinear stage of the simulations we observe secondary events obeying the same critical scaling with $S$, here calculated on the \emph{local}, much smaller lengths, leading to increasingly faster reconnection. These findings strongly support the idea that in a fully dynamic regime, as soon as current sheets develop, thin and reach this critical threshold in their aspect ratio, the tearing mode is able to trigger plasmoid formation and reconnection on the local (ideal) Alfv\'enic timescales, as required to explain the explosive flaring activity often observed in solar and astrophysical plasmas.
\end{abstract}


\section{Introduction}

Magnetic reconnection is thought to be the primary mechanism providing fast energy release, readily channeled into heat and particle acceleration, in astrophysical and laboratory magnetically dominated plasmas.  Within the macroscopic regime of resistive magnetohydrodynamics (MHD), however, classical reconnection models predict timescales, in highly conducting plasmas, which are too slow to explain bursty phenomena such as solar flares in the corona or tokamak disruptions. In particular, the Sweet-Parker model (hereafter SP) of two-dimensional, steady, incompressible reconnection \citep{Sweet:1958,Parker:1957} predicts a reconnection rate $M=v/c_A\sim S^{-1/2}$, where $v$ is the speed of the flow entering the reconnecting site, $c_A$ the Alfv\'en velocity based on the field far from the sheet and $S=Lc_A/\eta$ is the Lundquist number for a given magnetic diffusivity $\eta$ ($L$ is the current sheet length or breadth, identified with the macroscopic scale), which can be as high as $S\sim10^{12}$ in the solar corona, if simply due to collisional resistivity. Such a rate is way too slow to explain any of the impulsive phenomena described above.

As first demonstrated by means of 2D MHD simulations by \citet{Biskamp:1986} however, stationary reconnecting, SP-like sites become unstable once the Lundquist number exceeds a critical value of order $S\sim 10^4$,  and are subject to fast tearing modes and plasmoid formation when their aspect ratio $L/a$ becomes large enough, also increasing the local reconnection rate. Recent detailed linear analyses and simulations have confirmed these findings \citep{Loureiro:2007,Lapenta:2008,Samtaney:2009,Bhattacharjee:2009,Cassak:2009a,Huang:2010a,Uzdensky:2010}. In particular, the SP current sheet, of inverse aspect ratio $a/L\sim S^{-1/2}$, in the presence of the typical inflow/outflow pattern characterizing steady reconnection, was shown to be tearing unstable with growth rates $\gamma \tau_A\sim S^{1/4}$, where $\tau_A=L/c_A$. For a recent review on the latest theoretical works on 2D reconnection and secondary island (plasmoid) instabilities, from MHD to Hall regimes, see \citet{Cassak:2012}.

The existence of instabilities with growth rates scaling as a positive power of $S$ poses severe conceptual problems, since the ideal limit, corresponding to $S\to\infty$ would lead to infinitely fast instabilities, while it is well known that in ideal MHD reconnection is impossible. 

This issue was resolved by \citet{Pucci:2014} (PV hereafter), who studied the stability of current sheets with generic inverse aspect ratios $a/L\sim S^{-\alpha}$. The authors found showed that a critical exponent separates current sheets subject to slow instabilities, with growth rates scaling as a negative power of $S$, from the unphysical fast instabilities scaling as a positive power of $S$. Indeed, for $\alpha = 1/3$, they found the growth rate of the fastest reconnecting mode to become independent of Lundquist number. They therefore conjectured that current sheets should not collapse to aspect ratios greater than this critical value, at which point the instability, which they called the ``ideal'' tearing mode, leads to Lundquist-independent reconnection. For this aspect ratio, current sheets have a thickness up to 100 times larger than a typical SP reconnecting layer, and the instability developed X-points and plasmoids, thus preventing any collapse to the standard SP current sheet or any other steady configuration with $\alpha>1/3$. This novel ``ideal'' tearing instability is very fast, with an asymptotic growth rate $\gamma\tau_A \simeq 0.62$, and leads to the sudden formation of several plasmoids. In particular, PV found the relation $kL\sim S^{1/6}$, with $k$ the fastest growing wave-vector along the current sheet. 

In the present work we investigate numerically, by means of (compressible) resistive 2D-MHD simulations, the linear and nonlinear stages of the tearing mode for a current sheet at the critical thickness $a/L=S^{-1/3}$. Several initial configurations are tested, from the Harris sheet with fluid pressure balance, to the purely force-free case, with magnetic field rotation inside the current sheet, and also different values for the asymptotic plasma beta. 

The goal of this paper is, on the one hand, to retrieve all the known linear results and scalings, namely the expected instability dispersion relation and eigenmode structure (within the range of Lundquist numbers accessible to our simulations, that is up to $S=10^7$ to limit the computation time), on the other hand to explore the nonlinear regime of the ``ideal'' tearing instability for the first time. Our simulations provide further proof of the existence of such an instability, which is expected to set in during current sheet collapse arising in any turbulent scenario of plasma dynamics \citep{Loureiro:2009,Servidio:2009,Rappazzo:2013}.

The paper is structured as follows. In section~\ref{sect:setup} we describe the set of equations, the initial conditions, and our numerical setup. Section~\ref{sect:lin_an} is devoted to the numerical validation of the linear theory of PV. In section~\ref{sect:nonlin} we show the nonlinear results. Section~\ref{sect:concl} contains the discussions and conclusions.

\section{Numerical setup}
\label{sect:setup}

We integrate the compressible, resistive MHD equations numerically,
with the adiabatic index $\gamma = 5/3$, in the form
\be
 \frac{\partial \rho}{\partial t}  + \nabla\cdot ( \rho \mathbf{v} ) = 0,
\label{eq:mhd1}
\ee
\be
\frac{\partial  \mathbf{v}}{\partial t}  +  (\mathbf{v}\cdot \nabla)  \mathbf{v} = 
\frac{1}{\rho} \left[- \nabla p + (\nabla\times\mathbf{B}) \times\mathbf{B} \right],
\label{eq:mhd2}
\ee
\be
\frac{\partial T}{\partial t}  +  (\mathbf{v}\cdot \nabla) T = 
(\gamma - 1) \left[ - (\nabla\cdot \mathbf{v} ) T + \frac{1}{S} \frac{|\nabla\times\mathbf{B}|^2}{\rho}  \right],
\label{eq:mhd3}
\ee
\be
\frac{\partial  \mathbf{B}}{\partial t}  = \nabla \times (\mathbf{v}\times\mathbf{B} ) + \frac{1}{S} \nabla^2 \mathbf{B},
\label{eq:mhd4}
\ee
where $S$ is the Lundquist number defined above and other quantities retain their obvious meaning. Physical quantities are normalized using Alfv\'enic units, namely a characteristic length scale $L$, a characteristic density $\rho_0$, and a characteristic magnetic field strength $B_0/\sqrt{4\pi}$ (the background values measured far from the current sheet). Velocities are then expressed in terms of the Alfv\'en speed $c_A = B_0/\sqrt{4\pi\rho_0}$, time in terms of $\tau_A=L/c_A$, the fluid pressure in terms of $B_0^2/4\pi$. Note that we are using the energy equation~(\ref{eq:mhd3}) written for the normalized temperature $T=p/\rho$, where we use as a reference value $T_0 = (m/k_\mathrm{B})c_A^2$. With the given normalizations the Lundquist number is basically the inverse of the magnetic diffusivity, namely $S=c_A L/\eta\to \eta^{-1}$. 

The initial conditions at $t=0$ for our two-dimensional simulations of the tearing instability are different forms of the Harris current sheet configuration in which the equilibrium magnetic field varies only in the $x$ direction, reaching an asymptotic magnitude $B=1$ far from $x=0$, the plasma density is uniform $\rho=1$, and the pressure (and temperature) $p=T=\beta/2$ far from the current sheet, localized around $x=0$,  the asymptotic plasma beta being a given parameter. Two types of equilibrium are considered: in the first case, the classical Harris sheet, the field has only one component, the magnetic pressure gradient is balanced by a temperature enhancement in the current sheet itself (requiring a local plasma beta of order one, regardless of the value of  the parameter $\beta$), whereas in the second case the magnetic field is in a force-free equilibrium, $p$ and $T$ are taken constant everywhere, the condition $B^2=\mathrm{const}$ being preserved by the fact that the magnetic field rotates across the sheet, so that there is a non-vanishing component $B_z\neq 0$  inside the current sheet itself. Introducing a new parameter $\zeta$, the (normalized) maximum amplitude of the $z$ component of the magnetic field, both equilibria can be described writing:
\be
\mathbf{B} = \tanh\left(\frac{x}{a}\right) \hat{\mathbf{y}} + \zeta\,\mathrm{sech}\left(\frac{x}{a}\right) \hat{\mathbf{z}},
\label{eq:b0}
\ee
and for the fluid pressure is
\be
p=T=\frac{\beta}{2} + \frac{1-\zeta^2}{2}\,\mathrm{sech}^2\left(\frac{x}{a}\right),
\label{eq:p0}
\ee
with $\zeta = 0$ for the first case with $B_z=0$ in pressure equilibrium (PE hereafter), and $\zeta = 1$ for the force-free equilibrium (FFE hereafter).  Intermediate cases of mixed fluid/magnetic pressure equilibrium with $0\leq \zeta \leq 1$ can are also equilibria. 

With these normalizations, it is the thickness $a$ of the current sheet that defines the growth rate of the tearing instability: in the incompressible linear analysis by PV it has been shown that, when $a\sim S^{-1/3}$, for sufficiently high values of the Lundquist number (larger than $10^7-10^8$) the growth rate of the instability $\gamma$, measured in terms of the macroscopic Alfv\'enic time $\tau_A$, becomes independent of the magnetic diffusivity and of the order unity.  For the set of simulations shown below the current sheet thickness has been taken precisely $a=S^{-1/3}$, regardless of the equilibrium model chosen, i.e. the adopted values of $\beta$ and $\zeta$. 

The compressible, resistive MHD equations (\ref{eq:mhd1}-\ref{eq:mhd4}) are solved in a rectangular domain $[-L_x, L_x] \times [0, L_y]$ with resolution $N_x$ and $N_y$ respectively. In the $x$-direction, in order to resolve the steep gradients inside the current sheet using a reasonable number of grid points, we limit our domain to a few times the current sheet thickness, i.e. we set $L_x=20 a$: this is a good compromise between the high resolution required inside the current sheet and the need to have boundaries sufficiently far from the reconnecting region. Along the $y$-direction the length is chosen in order to resolve for the fastest growing modes of the instability (see below). 

The tearing instability is characterized by the exponential growth of modes with wavelength larger than the  current sheet thickness, that is with $ka < 1$ ($k$ being the mode wave-vector along $y$). The length $L_y$ along the current sheet is thus adapted to cover the range of unstable modes, namely we choose $L_y=m\lambda$, where $m$ is the number of wavelengths $\lambda=2\pi/k$ that we wish to simulate in our numerical box. Both lengths are chosen to decrease with $S$, 
\be
L_x=20a=20S^{-1/3}, \quad L_y = m \lambda = m \frac{2\pi}{ka}S^{-1/3},
\label{eq:LxLy}
\ee
where the value of $ka\sim S^{-1/6}$ is of order 0.1 for $S=10^7$, from the linear analysis by PV. The linear analysis in the next section is performed by taking sheet lengths for which only one mode ($m=1$), the most unstable one, is excited, whereas for the nonlinear simulations we will choose $L_y$ so that several unstable modes are independently excited (typically $m=4$), so to allow the subsequent mode-coupling and inverse cascade (i.e. the merging of plasmoids). 

In order to trigger the tearing instability, the equilibrium configuration is modified at $t=0$ with velocity perturbations of amplitude $\varepsilon\sim 10^{-3}$ (the rather large value speeds up the evolution) and wave-vector $\mathbf{k}=k\,\hat{\mathbf{y}}$, where $k$ is the same quantity appearing in equation~(\ref{eq:LxLy}), namely the wave-vector of the fastest growing mode selected for the analysis, as expected from the linear theory. Along the $x$-direction these velocities are concentrated at the current sheet location and vanish far from the current sheet. Moreover, the $v_x$ component is taken to be odd across the reconnection layer, whereas the $y$-component $v_y$, is obtained by imposing the perturbation velocity field to be incompressible. The analytical expressions for the perturbations are
\bea
v_x & = & \varepsilon \tanh\xi \, {\rm e}^{-\xi^2} \cos  (ky \! + \!  \varphi_k) , \\
v_y & = & \varepsilon (2\xi\tanh\xi \! - \!  \mathrm{sech}^2\xi\,) {\rm e}^{-\xi^2} S^{1/2} k^{-1} \sin (ky \! + \! \varphi_k),
\eea
where $\varphi_k$ is a random phase (for each mode $k$) and $\xi=x\,S^{1/2}$. 

The numerical simulations are performed by integrating equations (\ref{eq:mhd1}-\ref{eq:mhd4}) with an MHD code developed by our group. Along the current sheet, where periodicity is assumed, spatial integration is performed by using pseudo-spectral methods, while in the $x$-direction integration is performed by the use of a fourth-order scheme based on compact finite-differences \citep{Lele:1992}. The boundary conditions in the non-periodic direction are treated with the method of projected characteristics \citep{Poinsot:1992,Roe:1996,Del-Zanna:2001,Landi:2005}, here assuming non-reflecting boundary conditions. Time integration is performed using a third-order Runge-Kutta method. Details of the code are described in \cite{Landi:2005}.  

The resolution is adapted to the Lundquist number we use: for $S=10^5$ and $S=10^6$ we choose $N_x=1024$ and $N_y=128$, while for $S=10^7$ the number of cells in the $x$ direction is increased up to $N_x=2048$. In the periodic direction we use $N_y=128$  for the single-mode runs (in order to reduce the computational costs as many simulations are required to reproduce the instability dispersion relation curves), while we take $N_y=256$ in the nonlinear reference simulation. We have verified that this relatively low resolution along the periodic direction is adequate, due the extreme accuracy of Fourier methods and the rather smooth gradients observed in the $y$ direction. 
In spite of the relatively high values of S,  in addition to instability, the equilibrium diffuses on time-scales which although long compared to the instability, are still sufficient to affect linear evolution leading to slightly underestimate the growth rates of linear modes  \citep{Landi:2008}. To avoid this, in the single mode linear analysis described below, the diffusion term of the initial equilibrium is subtracted on the rhs of the induction equation at all times.


\section{Single mode simulations: linear analysis}
\label{sect:lin_an}

\begin{figure}[tb]
\begin{center} 
  \includegraphics[scale=0.43]{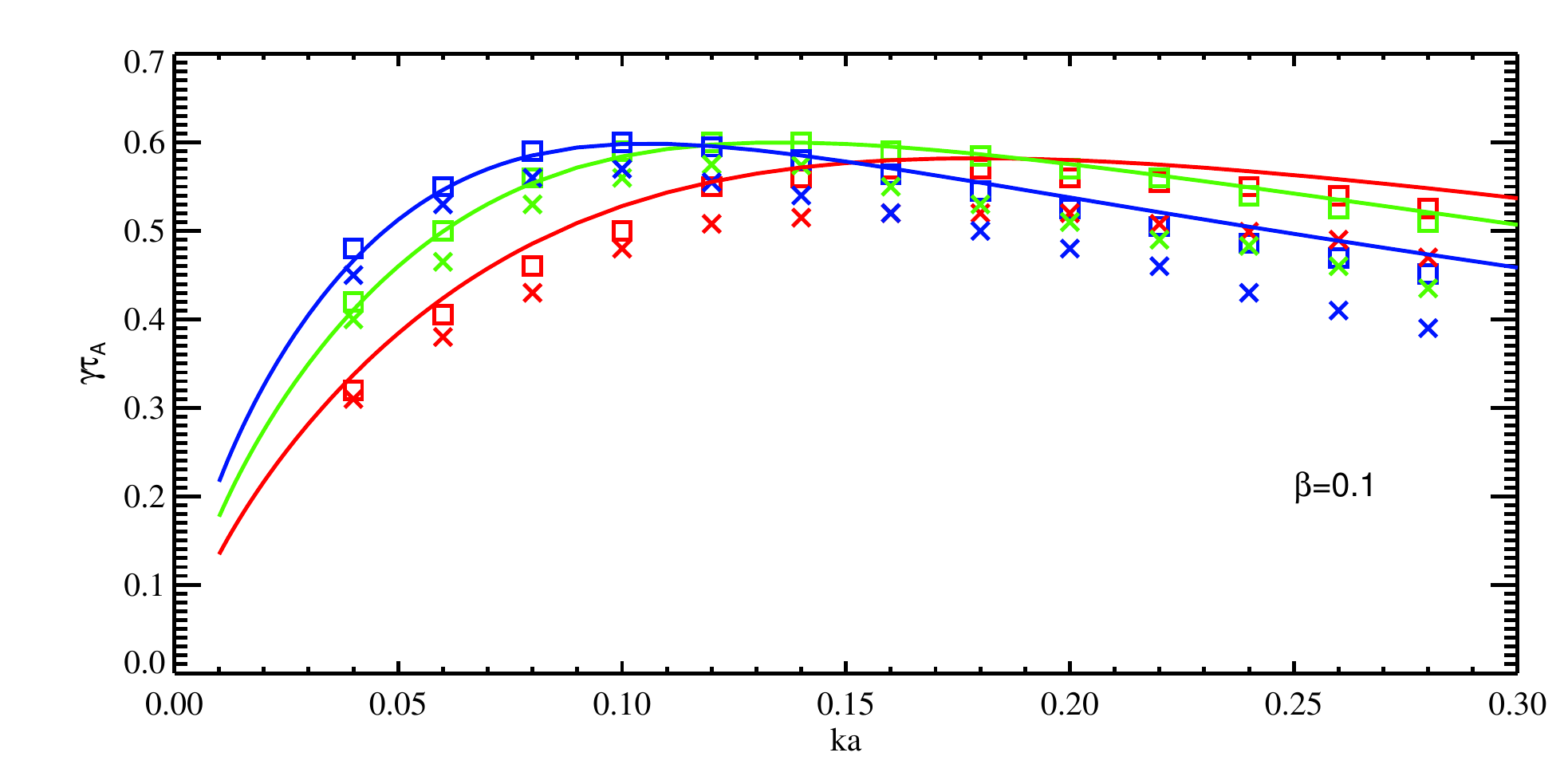}
  \includegraphics[scale=0.43]{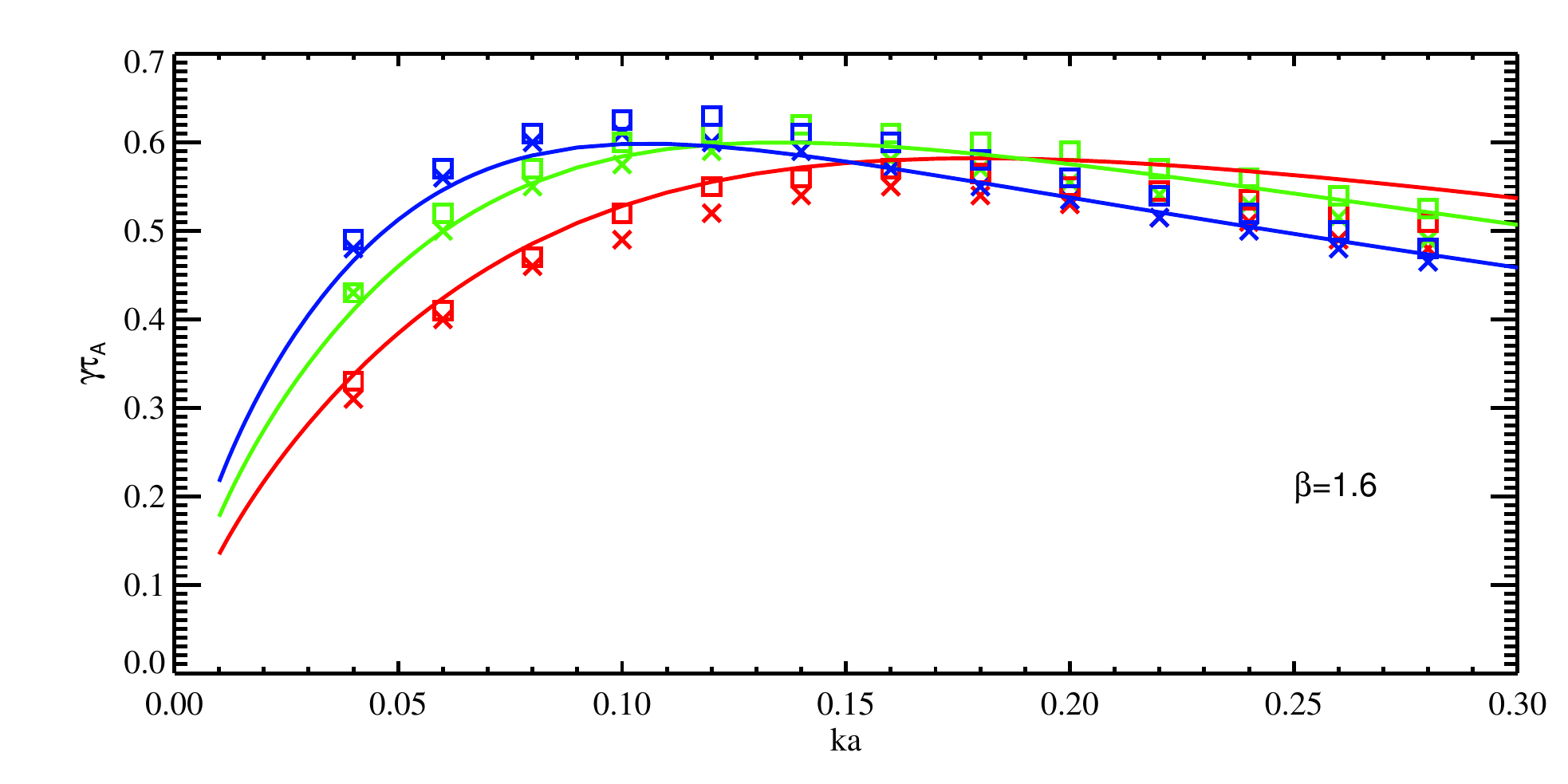} 
\end{center}
\caption{The instability dispersion relation (growth rate as a function of $k$, normalized against $a^{-1}=S^{1/3}$) for different values of the asymptotic beta (top panel: $\beta=0.1$; bottom panel: $\beta=1.6$), and Lundquist numbers (red color: $S=10^5$; green color $S=10^6$; blue color: $S=10^7$). Solid lines are the theoretical expectations, symbols are for numerical results (crosses: PE; squares: FFE).}
\label{fig:lindisp}
\end{figure}
\begin{figure*}[tb]
\center\includegraphics[width=130mm]{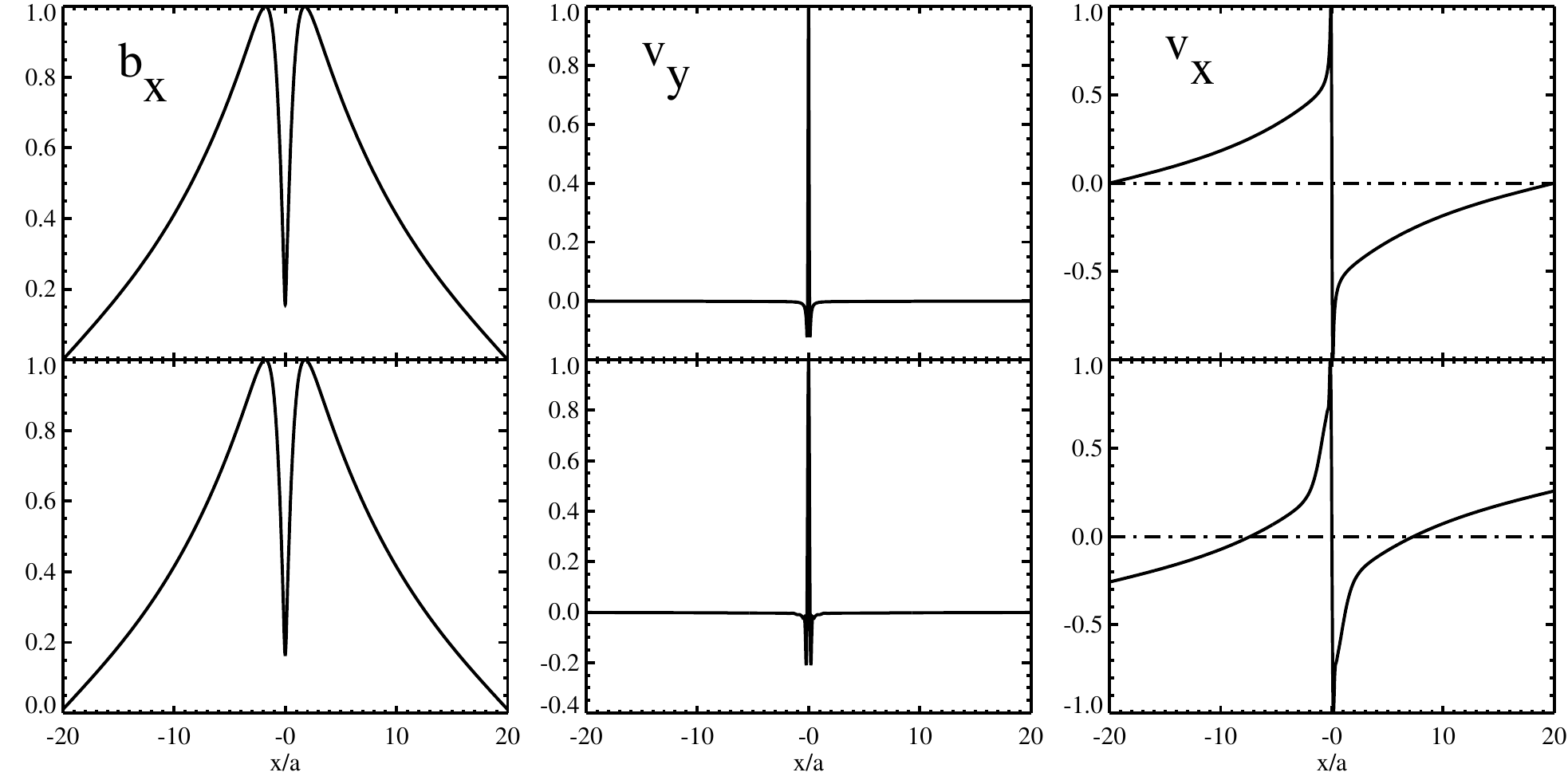}
\caption{Profiles across the current sheet of the tearing instability eigenmodes $b_x$ ($b_y$ is automatically determined by the solenoidal constraint), $v_y$, and $v_x$. Top panel: analytical linear theory; bottom panel: numerical results. All profiles are normalized to their maximum value.}
\label{fig:linfield}
\end{figure*}

A first set of simulations of the tearing instability in current sheets with $a=S^{-1/3}$ is performed to confirm the expected linear behavior, i.e. the scalings reported in \citet{Pucci:2014}, and in particular the instability dispersion relation as a function of the model parameters, here reported in figure~\ref{fig:lindisp} (solid lines). 

Even in the presence of the general equilibrium in equations (\ref{eq:b0}-\ref{eq:p0}) with $\zeta = {B_0}_z/B_0 \neq 0$, it is easy to show that the linear analysis of the instability is unchanged with respect to PV (in the incompressible limit $\rho=1$ and assuming perturbations in the $x-y$ plane alone with $\partial_z=0$). The governing equations for the tearing mode are still \citep{Furth:1963}
\bea
& & \gamma\, (v_x^{\prime\prime} - k^2 v_x) = ik [ {B_0}_y ( b_x^{\prime\prime} - k^2 b_x) - {B_0}_y^{\prime\prime} b_x], \\
& & \gamma\, b_x = ik {B_0}_y v_x + S^{-1}({b_x}^{\prime\prime} - k^2 b_x),
\eea
where we have assumed that perturbations are factorized as $\propto f(x)\exp(\gamma t +ky)$ and the prime denotes derivation with respect to $x$. Thus, even when $\zeta\neq 0$ there is no coupling of modes with the ${B_0}_z$ component, and the PV results for initial equilibria with $a=S^{-1/3}$ should remain unchanged. Moreover, no dependency on $\beta$ is expected, as neither this parameter enters the instability equations above, thus the theoretical dispersion relation curves only depend by the choice of $S$.

In spite of the expectations commented above for an incompressible situation, the results of a numerical simulation can deviate from the analytical case, due to the compressible regime, to differences in the treatment of boundary conditions (see below), and in general to discretization errors and other numerical approximations. Therefore, we choose to test the two limits of our initial equilibria for the current sheet, namely PE ($\zeta=0$) and FFE ($\zeta=1$). Moreover, we investigate both cases with $\beta<1$ ($\beta=0.1$) and $\beta>1$ ($\beta=1.6$), where $\beta$ is the asymptotic plasma beta in equation~(\ref{eq:p0}). Finally, three different values of the Lundquist number are tested here, namely $S=10^5$, $10^6$, and $10^7$, for a total of 12 sets of simulations of the linear phase of the tearing instability, with the aim of reproducing the expected dispersion relations numerically , as shown in figure~\ref{fig:lindisp}.

As anticipated in the previous section, for each value of $ka$ we vary $L_y$ while always selecting a single mode $m=1$. The growth rate of the instability is computed by measuring the $x$-averaged amplitude of the component $b_x$ of the perturbed magnetic field ($b_x=0$ at the initial time). In order to better compare with theoretical expectations, the PV eigenmode analysis have been redone here for a limited region across the current sheet of $-20\leq x/a \leq 20$, precisely as in our simulations. However, additional discrepancies are still expected, since the $v_x$ eigenmode is forced to vanish at the boundaries in the PV calculations, whereas in simulations we impose non-reflecting boundary conditions. 

The first thing to notice by inspecting the computed dispersion relations is that, as predicted by the classical linear theory, for each value of $S$ the curves have a maximum at a given $k$, the peak location decreasing in $k$ as $S$ increases. The growth rate of the instability (normalized to the inverse of the large-scale Alfv\'en time $\tau_A$) has peaks ranging from  $\gamma\approx 0.5$ for $S=10^5$ to $\gamma> 0.6$ for $S=10^7$. 


In general we find that the simulations with $\beta=1.6$ (bottom panel) are more precise in matching the analytical results than those with $\beta=0.1$ (top panel), since a large beta is a condition closer to incompressibility (formally corresponding to an infinite value for the sound speed). Moreover, we find that simulations of the FFE scenario (squares) yield higher and usually more accurate values of the growth rates as compared to those employing the PE settings (crosses): this is probably due to the fact that the purely force-free equilibrium leads to intrinsically less compressible fluctuations. Finally, rather large discrepancies are observed for small scales (large values of $ka$), especially in the PE case.

In figure~\ref{fig:linfield} we plot the profiles of the perturbations $b_x$ ($b_y$ is determined by $\nabla\cdot\mathbf{B}=0$), $v_y$ and $v_x$, all normalized to their respective maximum, across the current sheet in the $x$ direction. In the top panels we show the analytical results, that is the eigenmodes of the linear analysis (here the PV calculations have been recomputed by imposing $v_x=0$ for $x=\pm 20 a$), and in the lower panels we report the numerical solutions for a simulation in the FFE scenario with $S=10^7$ and $ka=0.10$, at a given time of the linear evolution of the instability. In order to recover the theoretical eigenmodes, velocity and magnetic field perturbations are shown with a $\pi/2$ shift in $ky$, as expected. Notice the steep gradients arising within the current sheet ($|x|\le a$), where a high resolution is needed to resolve the small scales developed during the instability evolution. 

As seen, the eigenmodes are very well reproduced: the magnetic field perturbations are identical to the analytical expected ones, while in the velocity perturbations the only major difference is, as anticipated, due to the non-reflecting free-outflow boundary conditions, that do not force $v_x=0$ at $x=\pm a$ and result in a slightly different profile even in the vicinity of the reconnecting region.

\section{Nonlinear simulations}
\label{sect:nonlin}

In the present section we investigate the nonlinear stages of the evolution of the tearing instability. Since we are interested in its late development, where interaction and merging of plasmoids is expected, we trigger the instability by selecting an initial spectrum of modes, rather than a single one as in the previous set of simulations, and we choose a maximum mode number $m_\mathrm{max}=10$. We also choose $L_y=1$ and $S=10^7$, so the modes with $$ka\simeq 0.029\,m; \quad m=1,10$$ are all excited. From the theoretical curves in the previous section we expect mainly a competition between modes $m=3$ and $m=4$ as the fast growing ones. As a reference run, we analyze the instability of a force-free equilibrium with constant temperature (FFE, $\zeta=1$), and we select the case with $\beta=1.6$. This combination was shown to provide a linear phase which is the closest to the analytical expectations (see figure~\ref{fig:lindisp}). The resolution employed for this run is $2048\times 256$, which is very high if one consider that the code employs high-order methods (compact finite-differences along $x$ and Fourier transforms along $y$, where periodical boundary conditions apply).

\begin{figure}[tb]
\hspace{5mm}
\center\includegraphics[width=80mm]{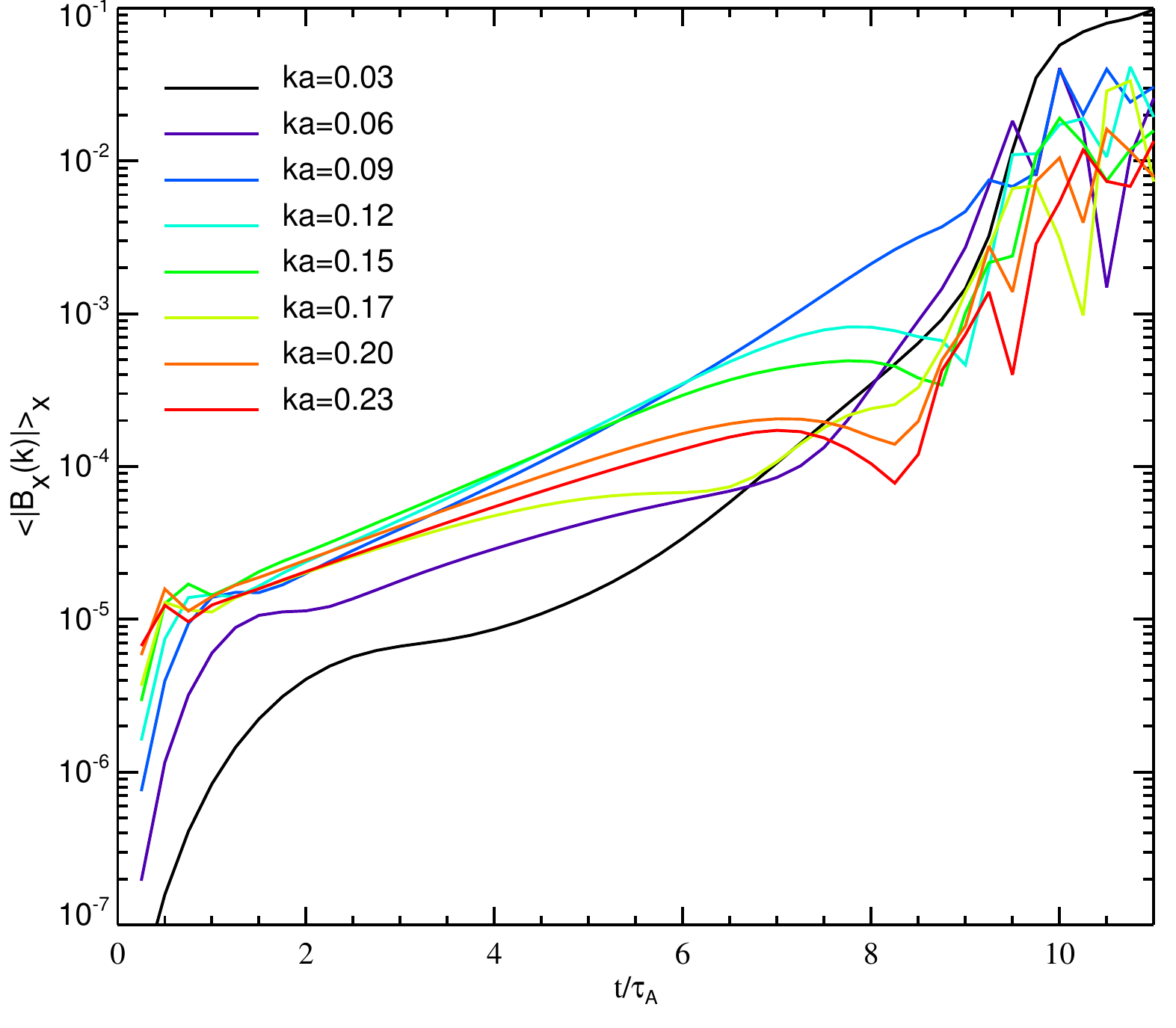}
\caption{
Nonlinear evolution of the tearing instability of a current sheet (FFE, $\zeta=1$, $\beta=1.6$) with $a=S^{-1/3}$. Growth of the excited perturbations (the first 8 modes, multiples of $ka = 0.03$) measured on the $x$ average of the $B_x$ component.}
\label{fig:ktime}
\end{figure}

\begin{figure*}[tb]
\center\includegraphics[width=18cm]{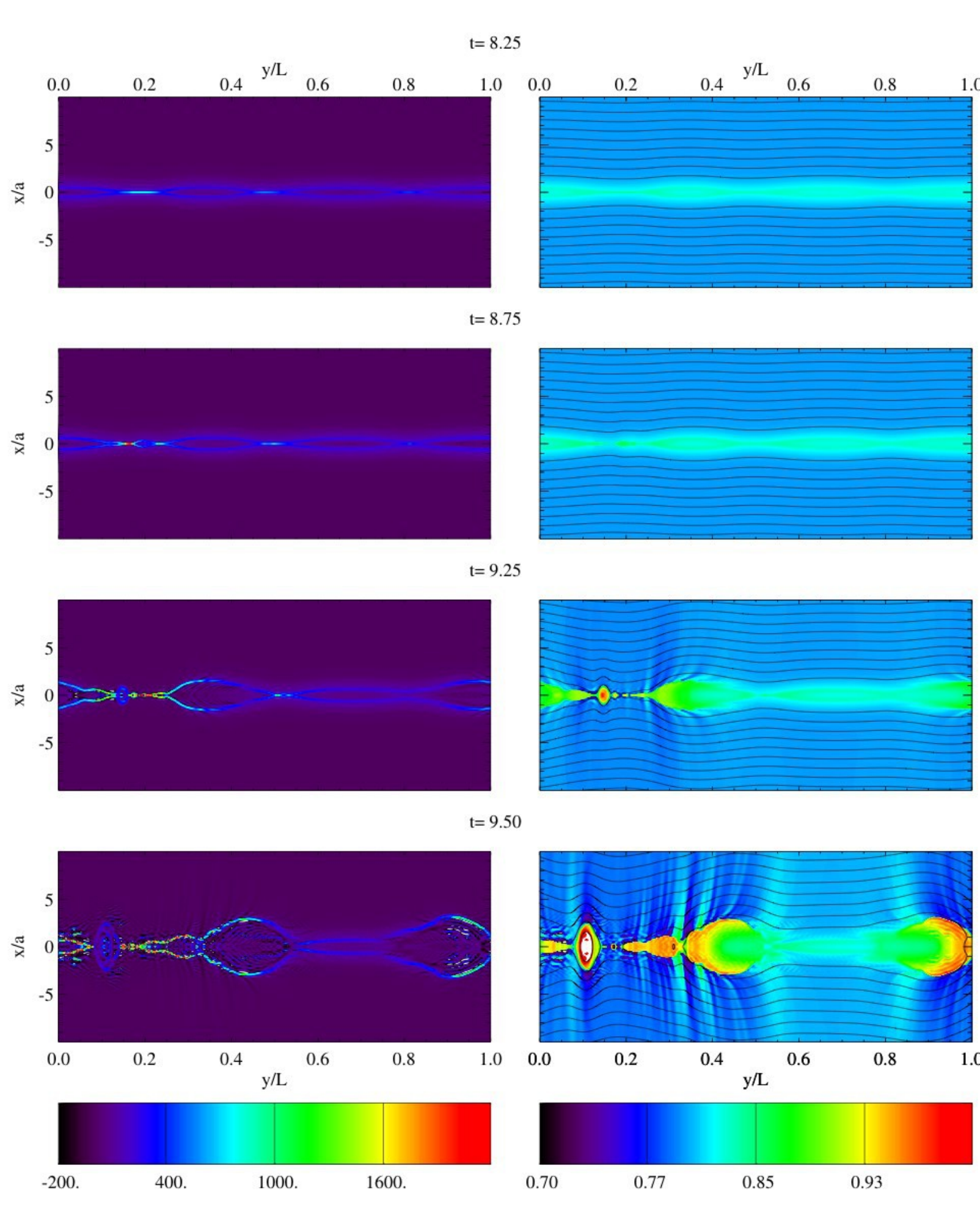}
\caption{
Nonlinear evolution of the tearing instability of a current sheet (FFE, $\zeta=1$, $\beta=1.6$) with $a=S^{-1/3}$. On the left panels we show the intensity of the $J_z$ electric current component, while on the right panels the magnetic fieldlines and the plasma temperature are displayed. Notice that the $x$ and $y$ scales are different and that we are zooming the inner region $|x|\leq 10a$, while the computation extends out to $|x|=20a$.}
\label{fig:nonlinear}
\end{figure*}

\begin{figure*}[tb]
\center
\includegraphics[width=55mm]{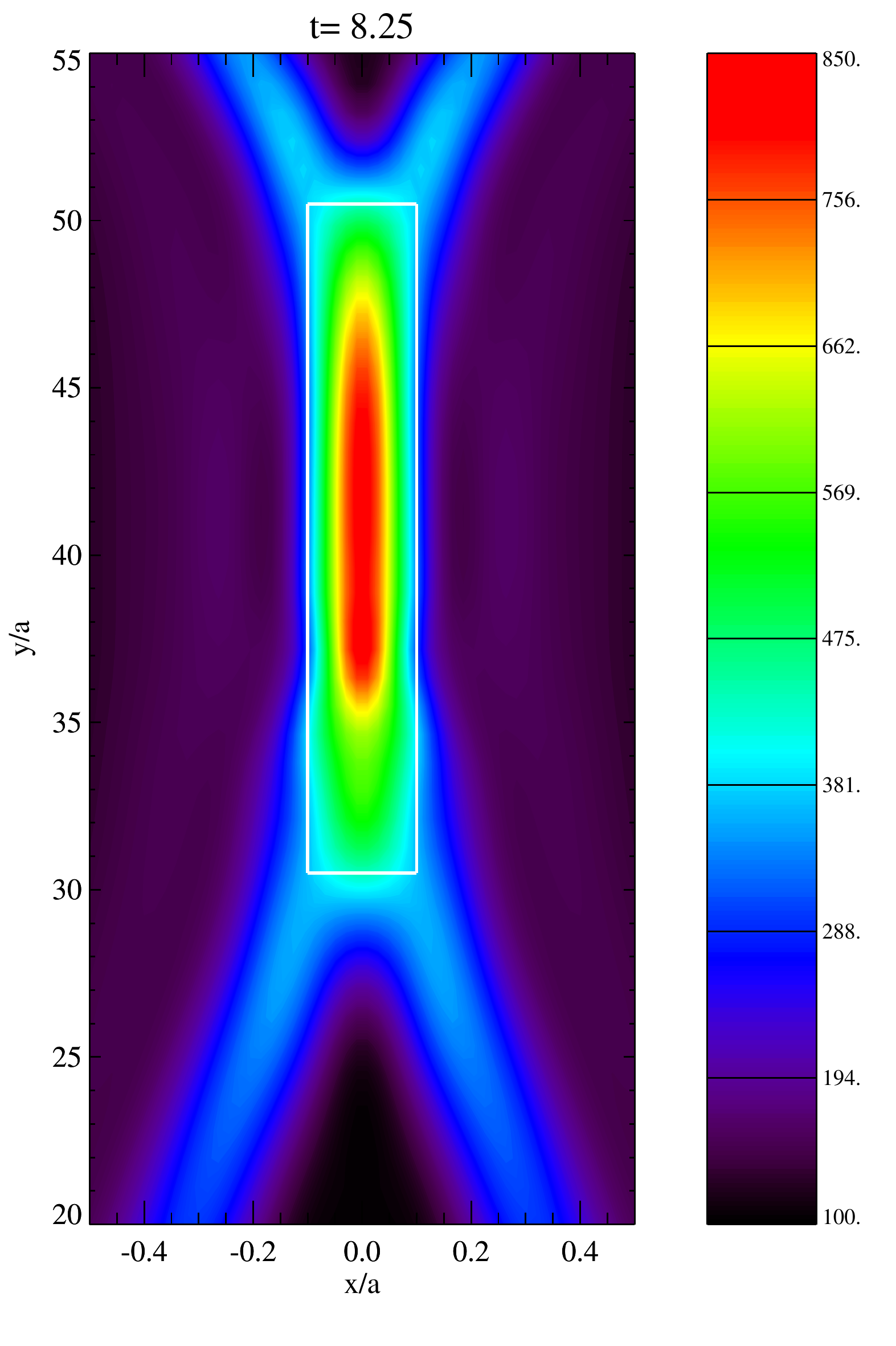} \quad
\includegraphics[width=55mm]{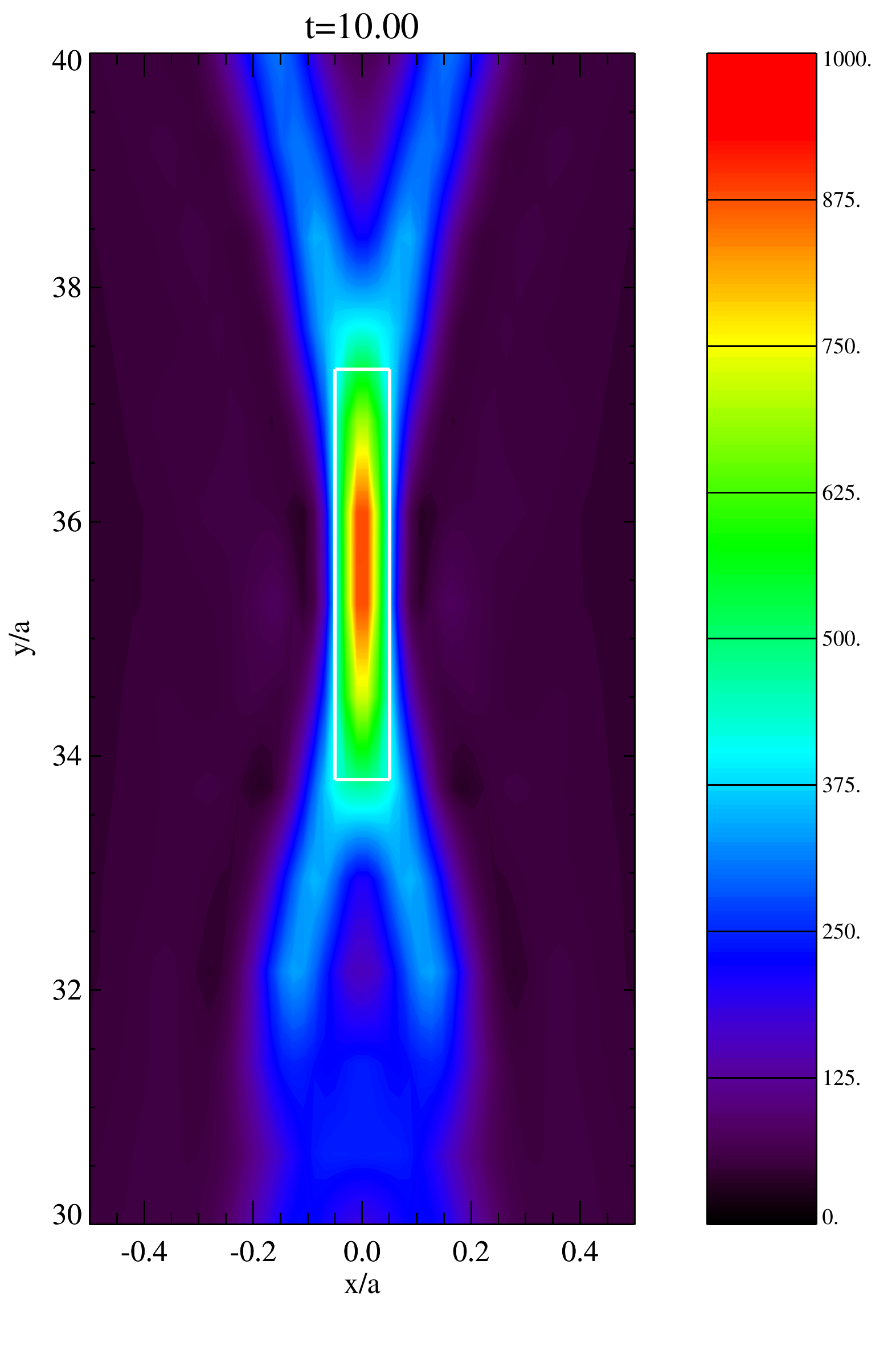} \quad
\includegraphics[width=55mm]{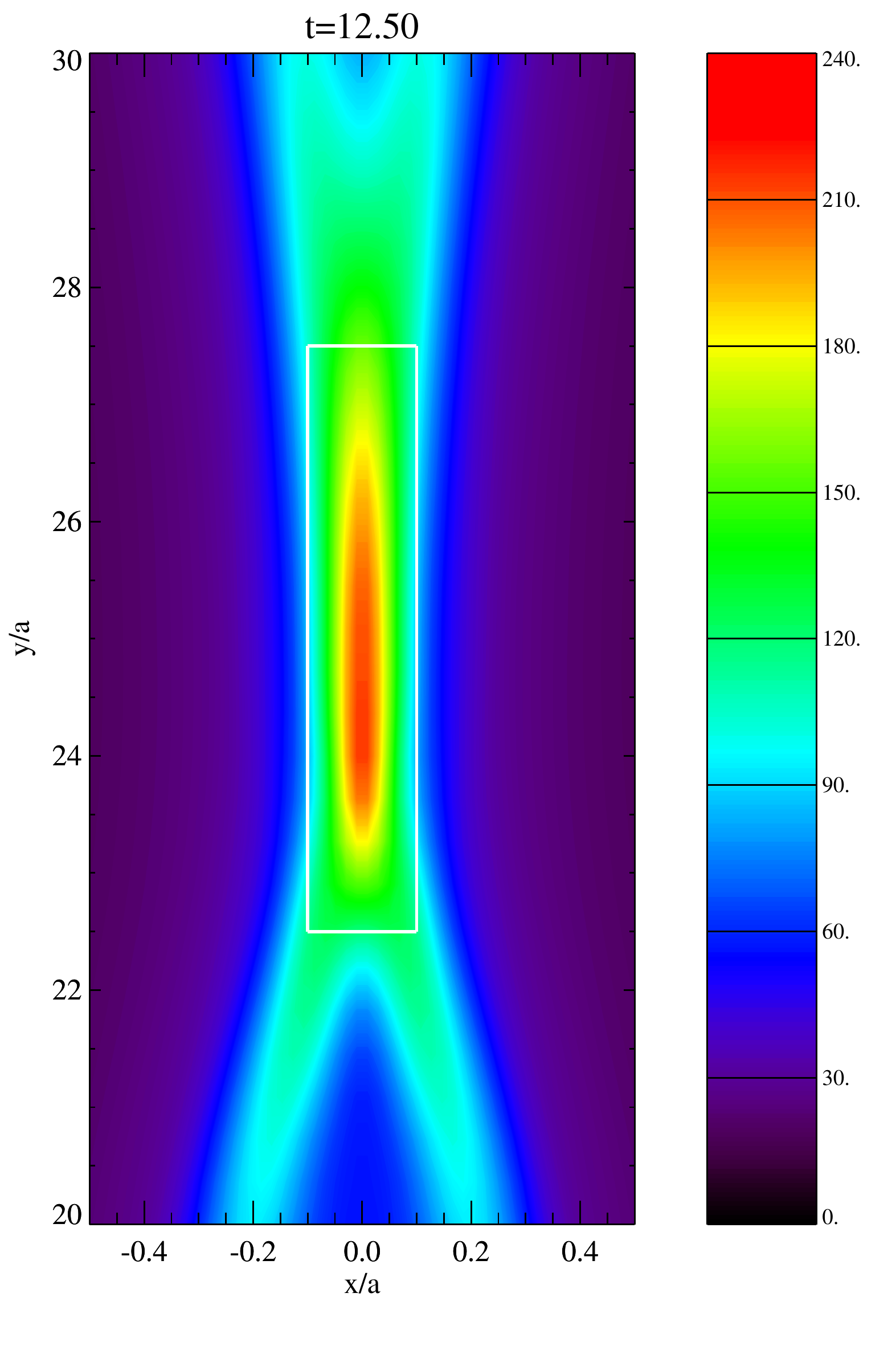}
\caption{
Zoom of the most prominent X-point reconnecting region, during the early nonlinear stage of the tearing instability. Colors refer to the strength of the $J_z$ component of the electric current. On the left panel we show the situation for the reference run with $S=10^7$, in the central panel the case $S=10^6$, in the right panel the case $S=10^5$.
}
\label{fig:zoom}
\end{figure*}

In figure~\ref{fig:ktime} we show the growth in time of the first 8 excited modes, based on a Fourier analysis along $y$ of the $B_x$ component averaged in the $x$ direction. Notice that towards the end of the linear phase ($t\simeq 5$) the dominant modes are $m=3$ ($ka\simeq 0.09$), with contributions from $m=4, 5$, as expected since their growth rates are similar. After $t\simeq6-7$  nonlinearities become important starts and energy in the longer wavelength modes increases due to the merging of smaller wavelength modes: as the $m=3$ mode is the most energetic one at this time, it keeps increasing its energy at the expense of higher m modes. At $t\simeq 9$ it has become dominant and the corresponding islands start to merge (see the full evolution described below): in the spectra this corresponds to a flattening of the  $m=3$ energy curve and to a strong increase first of the $m=2$ and finally the $m=1$ modes ($ka=0.03$), the largest available island in our y-periodic domain. This behavior is compatible with other 2D simulations of the tearing instability (see for example figure~3 in \citet{Landi:2008}).   

%

The complete evolution is shown in figure~\ref{fig:nonlinear}, where 2D snapshots for selected times are provided, of the region $|x/a|< 10$. Note that the figure has different scales in $x$ and $y$, the $x$ axis is normalized against $a$, the $y$ axis against $L$, with $L/a=S^{1/3}\simeq 215$). Panels on the left show the $J_z$ current component, whereas panels on the right show the distorted field lines superimposed on a map of temperature $T$. It is easy to recognize the end of the linear phase of the tearing instability (top row), with the dominant $m=3$ mode still clearly apparent. 

When the tearing instability growth is over, the nonlinear phase sets in leading to further reconnection events and eventually to island coalescence, departing from the dominant phase with $m=3$.  First, due to the attraction of current concentrations of the same sign, the X-points elongate and stretch along the $y$ direction \citep{Malara:1991,Malara:1992}. This process leads to a strong increase of the electric current concentrations and thus to the formation of new, elongated current sheets (see the second row, details of this phase will be discussed further on).

Beyond $t\simeq 9$ (third row) the evolution has become fully nonlinear and we clearly observe the process leading to the creation of a single, large magnetic island as arising from coalescence. The situation is very dynamic, especially near the major X-point where explosive expulsion of smaller and smaller islands is observed, typical of the plasmoid instability described in \cite{Loureiro:2007} and subsequent papers (in the spectra, as those shown in figure~\ref{fig:ktime}, the emergence of plasmoids is revealed by the oscillations of modes with  $m>1$). These islands then move towards the largest one, which is continually fed and thus further increases its size (approximately with a linear behaviour in time, until the boundaries are eventually reached) in a sort of inverse cascade eventually leading to the largest $m=1$ mode. Notice also the temperature enhancement at the reconnection sites. This long-term evolution is essentially determined by the imposed periodic boundary conditions along $y$.

At time $t=9.5$ (bottom row) both the current concentration and the plasma temperature around the X-point are so high that we need to saturate their values in order to retain an appropriate dynamical range in the color bar. The initial macroscopic current sheet is basically disrupted in a series of highly dynamical features. The current and temperature enhancements are stronger at the X-point and at the boundaries of the magnetic islands. Moreover, from the major reconnecting site we clearly see the production of magnetosonic waves, which propagate and soon steepen into shocks. 

At times $t>11$ the nonlinearities are so strong that the code, which is not conservative, cannot properly resolve the shock propagation and the simulation is interrupted. In addition, the largest island prevents the formation of new, smaller current sheets further feeding it, and the subsequent growth appears to proceed slowly, compatibly with the linear growth predicted by \citet{Rutherford:1973}.


Let us now investigate in more detail the situation right after the end of the linear phase of the tearing instability, when the islands coalescence is about to set in and the local current sheets have just formed and start to further evolve. In figure~\ref{fig:zoom} (left panel) we show the zoom around the X-point (which is just about to develop) near $y=0.2$ at $t=8.25$, the first row of figure~\ref{fig:nonlinear}. Here we display the electric current by using the same scale in both $x$ and $y$ directions, normalized against the macroscopic current sheet's width $a=S^{-1/3}\simeq 1/200 \simeq 5 \times 10^{-3}$, thus the position of the reconnection zone is now expressed as $y\simeq 40a$.

The local current sheet has just  formed as the result of a stretching process in the $y$ direction (and shrinking across the other direction), as a typical output of the nonlinear phase of the tearing instability. We are now in the phase in which, in the center of the current sheet, reconnection is about to take place, leading to a topology change in the magnetic structure and to the disruption of the current sheet itself, initially into two smaller strips. It is very interesting to measure the aspect ratio of this reconnection site. From the white box in the figure, defined by the rectangular region where the $J_z$ component has values which are roughly half those of the central peak, we estimate $L^*/a^*\simeq  200$, where we have used an asterisk to indicate the \emph{local values} and to differentiate with respect to the macroscopic ones (we also recall that in the whole paper we identify $a$ as the \emph{half} width of a current sheet). Additional simulations for $S=10^6$ and $S=10^5$, reported in the second and third panels, lead to local aspect ratios of $L^*/a^*\simeq 80$ and $L^*/a^*\simeq 50$, respectively.

If we now compare these numbers with ${S^*}^\alpha$, trying to find a value of $\alpha$ that fits the data best, it is easy to see that the value $\alpha=1/3$ is a very good guess. Here $S^*=(L^*/L)S$ is the local Lundquist number, which is obviously smaller than the macroscopic one, due to the much smaller length of the local current sheet (to be measured in each case). Therefore, based on our very limited data set, we derive the scaling
\be
L^*/a^* = k \, {S^*}^{1/3},
\label{eq:local}
\ee
where $k \simeq 2.1-2.3$, that is of order unity, as expected.

These findings are very important, in our opinion. For the first time we clearly see in simulations that, even in the nonlinear stages of the tearing instability, the new current sheets that form locally become unstable when the inverse aspect ratio of these structures reaches the critical threshold of $a^*/L^* \sim {S^*}^{-1/3}$, precisely the same limit found by PV for the fast reconnection of the initial, macroscopic current sheet. After that, a new ``ideal" tearing instability starts, with time occurring on a  \emph{faster} timescale $\tau_A^*=L^*/c_A$, since typically $L^*\ll L$. 

Furthermore, when smaller and smaller scales are produced nonlinearly, as observed at the time proceeds, each time the newly formed local current sheets elongate and reach their own critical value, that corresponding to equation~(\ref{eq:local}), faster and faster reconnection will arise producing a cascading, accelerating process: this, we believe, is the real nature of the plasmoid  (or super-tearing) instability.

In order to complete our numerical investigation, we have also performed nonlinear simulations for other settings, like the PE initial configuration and $\beta=0.1$. Some changes are seen during the linear stage, because of the slightly different growth rates, though the number of islands created by the fastest growing mode is always $m=3$. Coalescence to the $m=2$ and $m=1$ modes is invariably observed, as well as the rapid onset of the plasmoid instability. The runs with $\beta=0.1$ are less robust than the reference one with $\beta=1.6$, due to the stronger compressibility effects and faster growth of shock waves. We therefore deem that the nonlinear stages of the fast reconnection process described in this section have been captured correctly, and the crucial observation that even the secondary reconnection events occur when the local current sheets reach an inverse aspect ratio $\propto {S^*}^{-1/3}$, as for the initial settings, confirms that the physics of the ``ideal" tearing instability is indeed robust. 

\section{Conclusions}
\label{sect:concl}

In the present paper we have studied, by means of compressible, resistive MHD simulations, the linear and nonlinear stages of the tearing instability of a current sheet with inverse aspect ratio $a/L\sim S^{-1/3}$, where $S\gg 1$ is the Lundquist number measured on the macroscopic scale (the current sheet length) and the asymptotic Alfv\'en speed. Our results confirm the linear analysis of PV of the ``ideal'' tearing mode, while our nonlinear simulations show that the current sheet elongation and reconnection follows what appears to be a quasi-self-similar path, with subsequent collapse, current sheet thinning, elongation, destabilization starting from the X-points formed in the original sheet. As scales become smaller, and the effective Lundquist numbers decrease, the dynamical time-scales decrease, leading to explosive behavior. 

In particular, we have verified that the secondary reconnection events start their fast evolution precisely when the critical threshold $a/L\sim S^{-1/3}$ predicted in PV is replicated \emph{on the local scale}, according to equation~(\ref{eq:local}). We believe that this could be a universal behavior. Though a detailed renormalization-type analysis is beyond the scope of this paper, requiring the inclusion of finite inertial-length effects and therefore also the presence of guide fields etc., it seems clear that scaling phase diagrams for reconnection \citep[as provided for example by][]{Cassak:2012,Cassak:2013},
might end up being modified by the presence of this additional ``ideal" tearing regime and its kinetic generalizations currently under study. 

Notice that our analysis begins with a current sheet which has already reached the critical aspect ratio, but it is natural to speculate that because the tearing instability is incredibly slow, at large Lundquist numbers, this critical aspect ratio becomes \emph{de-facto} a trigger for explosive energy release. It would be interesting to perform large-Reynolds number simulations in which the current sheet thins, so that the aspect ratio increases with time: in such case the triggering of reconnection could be examined in greater detail. Even more realistic would be to perform simulations of turbulent reconnection \citep[see][and references therein]{Lazarian:2015}, though it would be hard to reach the accuracy needed to recognize our critical scaling in current sheets forming at all spatial scales.

Moreover, here only 2D MHD reconnection has been considered, while additional dynamics is expected in 3D simulations, with or without the presence of a guide field and/or shear flows, where the onset of the secondary plasmoid instability may be heavily modified by the interaction with other modes \citep{Onofri:2004,Landi:2008,Bettarini:2009,Landi:2012}.

An important limitation of our simulations is the use of only a resistive term in Ohm's law. \citet{Tenerani:2014} have considered the effects of viscosity on the ``ideal'' tearing scenario, showing that the main result is unchanged at Prandtl numbers of order unity, while greater critical aspect ratios, even close to the Sweet-Parker limit, are possible in more viscous regimes. On the other hand, for typical conditions of the solar corona, the smaller scales arising in non-linear evolution require the inclusion of further kinetic effects, such as the Hall term and electron pressure and inertial terms, in Ohm's law. While effects on the linear stability are presently being considered (Del Sarto et al., \emph{submitted}) it would be important for understanding the partitioning of energy between the bulk of the plasma and accelerated particles for fully kinetic simulations to be carried out.

The ``ideal'' tearing mode occurring on Alfv\'enic timescales naturally applies to the solar coronal plasma, where heating is supposed to be caused by the flaring activity at all scales (the Parker \emph{nanoflares} scenario) inside coronal loops \citep{Rappazzo:2007,Rappazzo:2008}. In addition, the same scenario could most probably be relevant also to relativistic pair plasmas \citep[see][and references therein]{Kagan:2015}, and in particular to the modeling of Pulsar Wind Nebulae \citep[e.g.][]{Porth:2014,Olmi:2014} in the light of the recent discovery of major gamma-ray flares observed in the Crab nebula \citep{Tavani:2011,Komissarov:2011,Cerutti:2012,Buhler:2014}.

\acknowledgements
M.V. was supported by the NASA Solar Probe Plus Observatory Scientist grant. The research leading to these results has received funding from the European Commissions Seventh Framework Programme (FP7/2007-2013) under the grant agreement SHOCK (project number 284515).

\bibliographystyle{apj}

\end{document}